\documentclass{article}
\usepackage{amsmath}

\def\be{\beta}

\def\be{\begin{equation}}
\def\ee{\end{equation}}

\newcommand{\de}{{\partial}}
\newcommand{\td}{{\tilde{d}}}
\newcommand{\tD}{{\tilde{D}}}

\newcommand{\Arth}{{\rm Arth}}
\newcommand{\artanh}{{\rm ar \, tanh }} 
\newcommand{\arcoth}{{\rm ar \, coth }}

\newcommand{\calM}{{\cal M}}

\newcommand{\Btheta}{{B_{\vartheta}}}

\newcommand{\unin}{\in \!\!\! / \,}
\newcommand{\BE}{\begin{equation}}
\newcommand{\EE}{\end{equation}}
\newcommand{\BEA}{\begin{eqnarray}}
\newcommand{\EEA}{\end{eqnarray}}
 
\begin{document}

\thispagestyle{empty}
 
\begin{flushright}
IFT-2/2003\\
  hep-th/0301212
\end{flushright}

\vspace{2cm}

\begin{center}
{\large\bf Nonsupersymmetric multibrane solutions}\\
\vspace{8mm}
{\bf Marcin P. Flak\footnote{
\ \ E-mail: Marcin.Flak@fuw.edu.pl},
Krzysztof A. Meissner\footnote{
\ \ E-mail: Krzysztof.Meissner@fuw.edu.pl} 
} 
\\ 
\vspace{4mm}
{\it
Institute of Theoretical Physics, Warsaw University\\ 
Ho\.za 69, 00-681 Warsaw, Poland\\
}
\end{center}

 
\vspace{1cm}
\begin{abstract}
Gravity coupled to an arbitrary number of antisymmetric tensors and scalar
fields in arbitrary space-time dimensions is studied in a context of
general, static, spherically symmetric solutions with many orthogonally
intersecting branes.  Neither supersymmetry nor harmonic gauge
is assumed.  It is shown that the system reduces to a Toda-like system
after an adequate redefinition of transverse radial coordinate $r$. 
Duality $r 
\rightarrow 1/r$ in the set of solutions is observed.
\end{abstract}    

\newpage
\section{Introduction}

Recently branes are subject of strong interest mainly because of their
crucial role in superstring theories where they can carry Ramond-Ramond
charges. However it is possible to view them from a classical point of
view as solutions of field equations in supergravity or more generally in
systems with gravity coupled to antisymmetric tensors and other fields.
Here we consider static, spherically symmetric solutions (see \cite{LP1,
Stelle, IM} and references therein). Most of solutions of this type were
previously found in so-called harmonic gauge (linearity condition) i.e.
assuming that supersymmetry is not entirely broken \cite{DGHR, DS}. There
were also attempts to relax this condition \cite{LPX, Tollsten}. In a case
of a system with single scalar and single antisymmetric tensor a complete
solution was presented in \cite{Zhou}. It was also observed that studying
multibrane systems is equivalent to solving a Toda-like system what turned
out to be a very helpful tool in derivation of many classes of exact
solutions (\cite{LPX, IK, CGM} and references in \cite{IM}). Such
equivalence however was proved with a postulate of Poincar\'e invariance 
in all directions parallel to at least one brane \cite{LPX} or in harmonic
gauge \cite{IK, CGM}.

In this paper we continue exploring this idea. We consider $D$-dimensional
gravity coupled to several antisymmetric tensors and scalars, assuming
that each antisymmetric tensor supports only one brane.  We do not
postulate Poincar\'e invariance in the whole subspace parallel to at least
one brane, but divide it into several smaller subspaces each uniquely
described as parallel to some of the branes and transversal to the others.
The geometry of each of the subspaces is governed by an independent factor
in the metric tensor. In the directions transversal to all branes
$SO(D-d)$ symmetry is assumed. We show that the classical equations of
motion are equivalent to a Toda-like system after an adequate redefinition
of radial coordinate: $r \rightarrow \vartheta(r)$, where $\vartheta(r)$
is in general not harmonic. A duality in the set of solutions is observed.
Each solution described by given values of parameters has a partner which
is numerically equal to it if a sign of $\vartheta$ is reversed (or
equivalently if $r \rightarrow 1/r$). In particular, the solution dual to
the supersymmetric (harmonic) is not supersymmetric (not harmonic).

\section{The model}

Consider a $D$ dimensional theory having in the classical limit the
following action:

 \BE 
\int_{\cal M} d^DX \sqrt{|\det g|} \left( 
  R - \frac{1}{2} \sum_{\alpha=1}^{N_\phi} \de_M 
\phi_\alpha \de^M \phi_\alpha
  - \sum_{i=1}^{N_A} \frac{\exp(\sum_{\alpha=1}^{N_\phi} 
a_{i\alpha} \phi_\alpha)}{2n_i!}
  F^i_{M_1\ldots M_{n_i}}F^{iM_1\ldots M_{n_i}} \right), 
\label{lagrangian_1} 
\EE
where $F^i$ are antisymmetric $n_i$-forms, $\phi^\alpha$ -- scalar fields,
$a_{i\alpha}$ -- constants, $\calM$ -- a manifold of dimension $D$ and
$(X^M)$ coordinates on it. For example the bosonic sector of most of
supergravity theories is well described by the above action if additional
 assumptions leading to cancellation of Chern-Simons term are made.
The equations of motions derived from (\ref{lagrangian_1}) are:

 \BEA 
R_{MN} &=& \frac{1}{2} \sum_\alpha \de_M \phi_\alpha \de_N \phi_\alpha 
   + \sum_i \frac{e^{\sum_\alpha a_{i\alpha} \phi_\alpha}}{2(n_i-1)!} 
S^i_{MN}, 
\label{eqm_11} \\
   0 &=& \nabla_M \left( e^{\sum_\alpha 
a_{i\alpha} \phi_\alpha}F^{iM R_1\ldots R_{n_i-1}} \right), 
\label{eqm_12} \\
   \nabla_M\nabla^M \phi_\alpha &=& \sum_i \frac{a_{i\alpha}}{2{n_i}!} 
e^{\sum_\beta a_{i\beta} \phi_\beta} 
    F^i_{R_1 \ldots R_{n_i}} F^{iR_1\ldots R_{n_i}}. 
\label{eqm_13} 
\EEA
with:
 \BE 
S^i_{MN}=F^i_{M R_1\ldots R_{n_i-1}} F^i_{N}{}^{R_1\ldots R_{n_i-1}}-
\frac{n_i-1}{n_i(D-2)}F^i_{R_1\ldots R_{n_i}} F^{iR_1\ldots R_{n_i}}g_{MN}. 
\EE
where $\sum_i$ and $\sum_\alpha$ are sums over all possible values of
$i=1,\ldots, N_A$ and $\alpha=1,\ldots, N_\phi$.

We search for a solution which allows $N_A$ orthogonally intersecting
elementary (electric) or solitonic (magnetic) branes, where each $V_i$ - a
worldvolume of the $i$-th brane is supported by a potential of the
adequate $F^i$ or $\ast F^i$. We define indices $I, J, \ldots$ running
through the set of all non-empty subsets of $\{1,\ldots,N_A\}$, subspaces
of $\calM$:

\BE 
\left. \begin{array}{ll} 
    V_I = \{0\} \cup \left(\left( \bigcap_{i\in I} V_i \right) \setminus 
\left( \sum_{j \unin I} V_j \right)\right), &
    V=\sum_i V_i=\oplus_I V_I \\    
    V_\emptyset = \calM/V, & \hat{V}_i=\sum_{I:i\unin I} V_I, 
  \end{array} \right. 
 \EE
(in other words, $V_I$ with $I=\{i_1,\ldots,i_k\}$ is a subspace spun by
vectors simultaneously parallel to all $V_{i_1}, \ldots, V_{i_k}$ and
transversal to all $V_{i_{k+1}}, \ldots, V_{i_{N_A}}$) and numbers of
 subspace's dimensions:

\BE 
d_I=\dim V_I, \qquad D_i=\dim V_i=\sum_{I:i\in I}d_I,
  \qquad \hat{D}_i = \dim \hat{V}_i=\sum_{I:i\unin I}d_I, 
\qquad d=\dim V=\sum_I d_I. 
\label{d_no} 
\EE
For particular brane configurations some of $V_I $ can be
zero-dimensional. (For example, if $V_i \subset V_j$ for any $i,j$, then
$V_{\{i\}}=\{0\}$.) We introduce also a mapping \hskip7pt $\tilde{}$
\hskip7pt defined by $\tilde{n}= D-n-2$.

Because all the branes are assumed to propagate in time and the solution
is trivial in case $V_\emptyset=\{0\}$ (all fields constant) therefore:

\BE  
1 \leq d_{\{1,\ldots,N_A\}} \leq D-1, \qquad -1 \leq 
\td \leq D-3, \qquad 0\leq d_I 
\leq D-2 \quad \mbox{for} 
\quad I \neq \{1,\ldots,N_A\}. 
\EE
Additionally it is postulated that:
\BE 
\td \neq 0. 
\EE
(This assumption can be omitted if an appropriate limiting procedure 
$\td\rightarrow 0$ is applied.)

We call by $(x^{\mu^i})$ the coordinates on $V_i$, by $(x^{\hat{\mu}^i})$
-- the coordinates on $\hat{V}_i$, by $(x^{\mu^I})$ -- on $V_I$, by
$(y^m)$ -- on $V_\emptyset$ and we use sum convention for all indices
enumerating coordinates but not for indices like $i$ or $I$. We introduce
also symbols $r=\sqrt{y^m y^n \delta_{mn}}$ and $f'=df/dr$ for any
function $f$.

We assume that all the fields depend nontrivially only on the radial 
coordinate (in the transverse space) $r$. For scalar fields it means:
\BE
\phi_\alpha(X) = \phi_\alpha (r).
\label{phi_anzatz}
\EE

Metric tensor is assumed to be divided into $N_g$ segments related to
$V_\emptyset$ and those $V_I$ which are at least one-dimensional ($2 \leq
N_g \leq 2^{N_A}$):

 \BE 
ds^2(X) = \sum_I e^{2A_I(r)} dx^{\mu^I} dx^{\nu^I} \eta_{\mu^I \nu^I}
  +  e^{2B(r)} \left( (dr)^2 + r^2 d\Omega^2 \right), \label{interval_1} 
\EE
where $d\Omega$ is the space interval of the $\td+1$ dimensional unit
sphere and $\eta_{\mu^I \nu^I}=\delta_{\mu^I \nu^I}$ if
$I\neq\{1,\ldots,N_A\}$. In the formula (\ref{interval_1}) and any
formulae below by $\sum_{I:r(I)}$ we denote a sum over those $I$ for which
$V_I$ are at least one-dimensional and which satisfy the restriction
$r(I)$.

For the antisymmetric tensor fields $F^i$, two cases should be distinguished.
If a brane is elementary, the only nonzero components of $F^i$ have a form:
 \BE 
F^i_{m \mu^i_1 \ldots \mu^i_{D_i}} (X) = \epsilon_{\mu^i_1 
\ldots \mu^i_{D_i}} \de_m \exp (C_i(r)) 
\label{el_Fansatz} 
\EE
when for a solitonic brane only:
\BE 
F^i_{\hat{\mu}^i_1 \ldots \hat{\mu}^i_{\hat{D}_i} m_1 \ldots m_{\td+1}} (X)
  = \epsilon_{\hat{\mu}^i_1 \ldots \hat{\mu}^i_{\hat{D}_i} m_1 \ldots m_{\td+1} n} 
\frac{\lambda_i y_n}{r^{\td+2}} 
\EE
do not vanish, where $\lambda_i$ is an arbitrary nonzero real constant.

Now it is possible to rewrite the equations of motion
(\ref{eqm_11}--\ref{eqm_13}) in terms of the scalar functions $A_I$, $B$,
$\phi_\alpha$ and (only for the elementary branes) $C_i$ introduced by
(\ref{interval_1}--\ref{el_Fansatz}) \cite{IM}: 

\BEA 
A''_I + A'_I \left( \sum_J d_J A'_J  + \td B' + \frac{\td+1}{r} \right) 
  &=& \frac{\sum_{i\in I} \tD_i(S'_i)^2 - \sum_{i\unin I} 
D_i(S'_i)^2}{2(D-2)}, 
\label{eqm_21} \\
  B'' + \td(B')^2 + \frac{2\td+1}{r}B' + (B' + \frac{1}{r}) \sum_I d_I A'_I
  &=& - \frac{\sum_i D_i(S'_i)^2}{2(D-2)}, \label{eqm_22} \\
  \phi''_\alpha + \phi'_\alpha \left( \sum_I d_I A'_I + 
\td B' + \frac{\td+1}{r} \right) 
  &=& -\frac{1}{2}\sum_i \varsigma_i a_{i\alpha} (S'_i)^2, 
\label{eqm_23} 
\EEA
 \BEA 
\td B'' - \td(B')^2 - \frac{\td}{r}B' + \sum_I d_I 
\left(A''_I -\frac{1}{r}A'_I-2A'_I B'+ (A'_I)^2 \right)
  + \frac{1}{2}\sum_\alpha (\phi'_\alpha)^2 &=& 
\frac{1}{2}\sum_i (S'_i)^2, \label{eqm_24} \\
  C''_i + C'_i \left( C'_i - \sum_{I:i\in I} d_I A'_I + 
\sum_{I:i \unin I} d_I A'_I + \td B' + 
  \sum_\alpha a_{i\alpha} \phi'_\alpha + \frac{\td+1}{r} \right) 
&=& 0, 
\label{eqm_25} 
\EEA
where:
\BEA 
\varsigma_i &=& \left\{ \begin{array}{ll} 
   +1 & \mbox{(elem.),} \\
   -1 & \mbox{(solit.),} \end{array} \right. \\
  S'_i &=& \left\{ \begin{array}{ll} 
   \exp(\frac{1}{2}\sum_\alpha a_{i\alpha}\phi_\alpha-
\sum_{I:i\in I} d_I A_I)(e^{C_i})' & \mbox{(elem.),} \\
    & \\
   \exp(\frac{1}{2}\sum_\alpha a_{i\alpha}\phi_\alpha-\sum_{I:i\unin I} 
   d_I A_I-\td B)
    \frac{\lambda_i}{r^{\td+1}}& \mbox{(solit.).} \end{array} \right. 
\label{S_1} 
\EEA

\section{Harmonicity, supersymmetry and $\vartheta$ coordinate}

Assuming that $\td \neq 0$, define:
\BE 
\chi = \sum_I d_I A_I + \td B, 
\label{lincon_1} 
\EE
Relation $\chi'=0$ is usually called linearity condition or harmonic gauge
because it leads to a solution expressed in terms of harmonic functions
on $V_\emptyset$ \cite{IM, Gibbons}.
In case of supergravity theories it is necessary but not sufficient 
for preserving supersymmetry \cite{DGHR, DS}.
Here we do not make any a priori assumption on $\chi$, so the results 
presented below remain valid in more general classes of non-supersymmetric 
solutions and solutions not governed by harmonic functions.

Summing (\ref{eqm_21}--\ref{eqm_22}) one can see that $\chi$ has to 
satisfy the following equation:
\BE 
\chi''+(\chi')^2+\frac{2\td +1}{r}\chi' = 0, 
\EE
which can be solved as:
\BE 
\chi(r)=\ln \left| \frac{c_\chi-1/r^{2\td}}{c_\chi-c_0} \right| 
+\epsilon_\chi, 
\label{chi1} 
\EE
where $\epsilon_\chi$ is an arbitrary real constant and $c_\chi$ takes real 
as well as infinite values. 
Since $\lim_{c_\chi\rightarrow +\infty} \chi = 
\lim_{c_\chi\rightarrow -\infty} \chi$, points $c_\chi=+\infty$
and $c_\chi=-\infty$ in the parameter space can be identified.
In the discussion below we choose $c_0=1$, but it can be 
generalised to arbitrary $c_0$.

Let us introduce new parameters: $R \in [0,+\infty]$ and 
$s_\chi \in \{-1,+1\}$, such that:
\BE 
s_\chi R^{2\td}=1/c_\chi 
\EE
(if $R=0$ or $R=\infty$ both possible signs of $s_\chi$ describe the same 
point in the parameter space) and define a function:
\BEA 
\vartheta(r) &=& \left\{ \begin{array}{ll}
\frac{1}{2\td}\left(\frac{1}{R^\td}+R^\td \right)
\left(\arctan((\frac{R}{r})^\td) - \arctan(R^\td) \right),
& s_\chi=-1, \\
\frac{1}{2\td}\left|\frac{1}{R^\td}-R^\td \right|
\left(\Arth((\frac{R}{r})^\td) - \Arth(R^\td) \right),    
& s_\chi=+1, 
\end{array} \right. 
\label{th_1} 
\EEA
 where:
\BE 
\Arth(x) = \left\{ \begin{array}{ll} 
  \artanh (x), & \mbox{   for   } |x|<1, \\
 -\arcoth (x), & \mbox{   for   } |x|>1.
 \end{array} \right. 
\label{Arth} 
\EE
Formula (\ref{th_1}) exhibits the following duality:
\BE 
\vartheta(r;R,s_\chi)=-\vartheta(1/r;1/R,s_\chi), 
\label{sym_1} 
\EE
what gives a relation between $\vartheta$ defined for different 
parameter values.
$R=0$ is equivalent to harmonic gauge and then and only then
$\vartheta= \frac{1}{2\td}\left(\frac{1}{r^\td} - 1 \right)$ what is 
harmonic function of $r$.
So, the parameter $R$ (or $c_\chi$) can be treated as a measure how distant 
is a given case from the harmonic one.
$R=\infty$, is a partner of $R=0$ under (\ref{sym_1}) and then
$\vartheta= -\frac{1}{2\td}\left(r^\td - 1 \right)$. 

Function $\vartheta$ can be understood as a space coordinate instead of $r$ 
and the coordinate change is singular only at $r=R$, $s_\chi=+1$.
The space-time interval expressed in terms of $\vartheta$ is:
\BE 
ds^2(\vartheta)= \sum_I e^{2A_I(\vartheta)} dx^{\mu^I} dx_{\mu^I} + 
 e^{2\Btheta(\vartheta)} \left( d\vartheta^2 + 
 \rho(\vartheta)^2 d\Omega^2 \right), 
\label{interval_2} 
\EE
where 
\BEA 
\exp(\Btheta(\vartheta)) &=& 
\left( \frac12 \exp(-\sum_I d_I A_I(\vartheta) +\epsilon_\chi) \right)^{1/\td} 
  \rho(\vartheta)^{-(1+1/\td)}, \label{btheta} \\
  \rho(\vartheta) &=& \left\{ \begin{array}{ll}
    \left| \frac{1}{4} ((1/R)^{\td}+R^{\td}) 
    \sin (\frac{4\td\vartheta}{(1/R)^{\td}+R^{\td}}+2\arctan R^\td)\right|,
   & s_\chi=-1,\\
    \left| \frac{1}{4} ((1/R)^{\td}-R^{\td}) 
    \sinh(\frac{4\td\vartheta}{\left|(1/R)^{\td}-R^{\td}\right|}+2\Arth R^\td
    )\right|,
   & s_\chi=+1 
 \end{array} \right. 
\label{rho_1} 
\EEA
and for $\rho$ one has:
\BE 
\rho(\vartheta;R,s_\chi)=\rho(-\vartheta;1/R,s_\chi) 
\label{sym_2}. 
\EE 
Note, that (\ref{interval_2}) is well defined even for such values of 
$\vartheta$ which cannot be related to any $r$ by (\ref{th_1}).
In other words, $\vartheta$ covers wider area of space-time than $r$. 
We discuss some aspects of the fact in the last section of the paper.

The equations (\ref{eqm_21}--\ref{eqm_25}) can be translated to:
\BEA 
  \ddot{A}_I &=& \frac{\sum_{i\in I} \tD_i(\dot{S}_i)^2 - 
  \sum_{i\unin I} D_i(\dot{S}_i)^2}{2(D-2)}, \label{eqm_31} \\
  \ddot{\phi}_\alpha &=& 
  -\frac{1}{2}\sum_i \varsigma_i a_{i\alpha} (\dot{S}_i)^2, 
  \label{eqm_32}\\
  0 &=& \ddot{S}_i + \left(\sum_\alpha a_{i\alpha} \dot{\phi}_\alpha - 
  \sum_{I:i\in I} d_I \dot{A}_I \right)\dot{S}_i,
  \quad \mbox{(only elem.)} 
  \label{eqm_34} 
\EEA
\BE 
\frac{1}{\td} \left( \sum_I d_I \dot{A}_I \right)^2 + 
\sum_I \left( d_I (\dot{A}_I)^2 \right) 
+ \frac{1}{2}\sum_\alpha(\dot{\phi}_\alpha)^2+\Lambda_\chi = 
\frac{1}{2}\sum_i (\dot{S}_i)^2, \label{eqm_33} 
\EE
where:
\BEA 
\dot{S}_i &=& \left\{ \begin{array}{ll} 
\exp(\frac{1}{2}\sum_\alpha a_{i\alpha}\phi_\alpha-
\sum_{I:i\in I} d_I A_I)(e^{C_i})\dot{} & \mbox{(elem.),} \\
& \\
-2\lambda_i e^{-\epsilon_\chi}
\exp(\frac{1}{2}\sum_\alpha a_{i\alpha}\phi_\alpha+
\sum_{I:i\in I} d_I A_I)& \mbox{(solit.),} 
\end{array} \right. 
\label{S_2} \\
\Lambda_\chi &=& -\frac{16\td(\td+1)c_\chi}{(c_\chi-1)^2} 
\label{kosmo_1}
\EEA
and the "dots" describe derivatives with respect to $\vartheta$.

The system (\ref{eqm_31}--\ref{kosmo_1}) together with 
(\ref{btheta}--\ref{rho_1}) carries complete information originaly
contained in (\ref{eqm_21}--\ref{eqm_25}, \ref{S_1}).
Since (\ref{eqm_21}--\ref{eqm_25}, \ref{S_1}) drastically simplifies when 
harmonic gauge is imposed, it is interesting what
happens to (\ref{btheta}--\ref{rho_1}, \ref{eqm_31}--\ref{kosmo_1}) 
in analogous situation.
If one treats $\vartheta$ as a fundamental coordinate then all dependence 
of the solution on parameter $R$ (so also all differences between the 
harmonic and general cases) enters only in two places.
In $\rho$ function (\ref{rho_1}) which
influences a form of the metric on $V_\emptyset$ and in 
$\Lambda_\chi$ constant (\ref{kosmo_1}) appearing in (\ref{eqm_33}).
However (\ref{eqm_33}) is not a dynamic equation but rather a constraint 
decreasing by one a number of integration constants.
What is more, if $R\neq 1$ two different $R$ gives the same value of 
$\Lambda_\chi$.
In particular, for both $R=0$ and $R=\infty$ one has $\Lambda_\chi=0$ and
$\rho(\vartheta)=|\td \vartheta|$.

\section{Toda-like system}

Define:
\BE 
|\omega_i| = \exp \left( \frac12 \sum_\alpha \varsigma_i a_{i\alpha}
\phi_\alpha-\sum_{I:i\in I} d_I A_I \right). 
\EE
With these functions one can find that (\ref{eqm_34}) leads to:

\BE 
\dot{S}_i = p_i/\omega_i, 
\label{S_3} 
\EE
where $p_i$ are nonzero real integration constants.
Simultaneously (\ref{S_2}) for solitonic branes gives 
$\dot{S}_i=\frac{-2\lambda_i e^{-\epsilon_\chi}}{\omega_i}$,
so, after identification $p_i=-2\lambda_i e^{-\epsilon_\chi}$, relation 
(\ref{S_3}) is valid for elementary as well as for solitonic branes.

It can be checked from (\ref{eqm_31}--\ref{eqm_32}) that $\omega_i$ have to 
satisfy the following system of equations: 

\BE 
\frac{d^2}{d\vartheta^2} (\ln |\omega_i|) = 
- \sum_j \Delta_{ij} \frac{p_j^2}{4\omega_j^2}, 
\label{toda_1} 
\EE
where elements of $\Delta$ matrix are:

\BE 
\Delta_{ij}= \frac{2}{D-2} \left(
\sum_{\bar{I}:i,j\in \bar{I}} d_{\bar{I}} 
\sum_{\bar{J}:i,j\unin \bar{J}} d_{\bar{J}} - 
\sum_{\bar{I}:i \in \bar{I},j \unin \bar{I}} d_{\bar{I}} 
\sum_{\bar{J}:i \unin \bar{J},j \in \bar{J} }d_{\bar{J}}  \right)
+ \sum_\alpha \varsigma_i a_{i\alpha} \varsigma_j a_{j\alpha}, 
\EE
indices $\bar{I}, \bar{J}$ run through all values allowed for $I, J$ and 
additionally $\emptyset$, and by $d_\emptyset$ is understood $\td$ 
(but not $\dim V_\emptyset$).
The diagonal elements of $\Delta$ are:

\BE 
\Delta_{ii}=\frac{2 D_i \tD_i}{D-2} + \sum_\alpha a^2_{i\alpha}, 
\EE
and the non-diagonal ones are bounded by:

\BE 
\Delta_{ij} \leq \frac12 (\Delta_{ii}+\Delta_{jj}). 
\EE
The equations (\ref{toda_1}) are equivalent to a Toda-like system.
After substituting $x_i=2 \ln( \frac{p_i \sqrt{\Delta_{ii}}}{2 \omega_i})$ 
they transform to:

\BE 
\ddot{x}_i=-\sum_j K_{ij} \exp(x_j). 
\label{Toda_2} 
\EE
where $K_{ij}=\frac{-2\Delta_{ij}}{\Delta{jj}}$.
The original Toda (molecule) system is defined by (\ref{Toda_2}) with $K$ 
being a Cartan matrix \cite{AC}.
But in our case $K$ is not in general a Cartan matrix.

If $\det(\Delta) \neq 0$ then it is possible to express all functions
$A_I$, $\Btheta$ (\ref{interval_2}), $\phi_\alpha$ (\ref{phi_anzatz}),
$C_i$ (\ref{el_Fansatz}) in terms of $\omega_i$:

\BEA
  \exp(A_I(\vartheta)) &=& E_I \left( \prod_i 
|\omega_i(\vartheta)|^{\gamma^i_I} \right) 
\exp(c_I\vartheta), \label{sol_11} \\
  \exp(\Btheta(\vartheta)) &=& E_B \left( 
\prod_i |\omega_i(\vartheta)|^{\gamma^i_B} \right) \exp(c_B\vartheta) 
   \rho(\vartheta)^{-(1+1/\td)} , \\
  \exp(\phi_\alpha(\vartheta)) &=& 
   E_\alpha \left( \prod_i |\omega_i(\vartheta)|^{\gamma^i_\alpha} 
\right)\exp(c_\alpha\vartheta),  
\label{sol_13} \\
  \frac{d}{d\vartheta} \exp(C_i(\vartheta))&=& p_i \left|
\omega_i(\vartheta)\right|^{-2}, 
\label{sol_14} 
\EEA
where $\gamma^i_I$, $\gamma^i_B$ and $\gamma^i_\alpha$ have to satisfy:

\BE 
\frac{D-2}{2} \sum_i \Delta_{ij} \gamma^i_I = \left\{ \begin{array}{ll} 
     -\tD_j & \mbox{if    }  j \in I, \\
        D_j & \mbox{if    }  j \in \!\!\!\!\! / \;I,
  \end{array} \right. \qquad
  \sum_i \Delta_{ij} \gamma^i_\alpha = 2 a_{j \alpha}, 
\qquad \gamma^i_B = - \frac{1}{\td} \sum_I d_I \gamma^i_I 
\EE
and values of real constants $c_I$, $c_B$, $c_\alpha$ and positive constants 
$E_I$, $E_B$, $E_\alpha$ are restricted by:

\BE 
\left. \begin{array}{rclcrcl} 
  0 &=& \frac12 \sum_\alpha \varsigma_i a_{i\alpha} c_\alpha - 
 \sum_{I:i\in I} d_I c_I, & \qquad &
  \prod_{I:i\in I}E_I^{d_I} &=& \prod_\alpha E_\alpha^{\frac12\varsigma_i 
 a_{i\alpha}}, \\
  0 &=& \sum_I d_I c_I +\td c_B, & \qquad & 
  \frac{E_\chi}{2}&=& \left( \prod_I E_I^{d_I} \right) E_B^\td. 
\end{array} \right. 
\EE
So, the problem of finding brane solution in gravity coupled to an arbitrary 
number of antisymmetric tensors and scalar fields without assumption of 
harmonic gauge can be reduced to solving a Toda-like system (\ref{toda_1}) 
with a condition derived from (\ref{eqm_33}):

\BE 
\sum_i N_i \frac{d}{d\vartheta}(\ln|\omega_i|) + \frac12 (\Lambda_\chi + 
\Lambda_c)  = \sum_i \frac{p_i^2}{4 \omega_i^2}, 
\label{eq_en} 
\EE
where $\Lambda_c$ and $N_i(\vartheta)$ are defined by:
 \BE 
\Lambda_c = \sum_I d_I c_I^2 + \td c_B^2 + 
\frac12 \sum_\alpha c_\alpha^2, \qquad  
  \sum_j \Delta_{ij} N_j = \frac{d}{d\vartheta}(\ln|\omega_i|). 
\EE

\section{An example of a solution}

Consider an example when matrix $\Delta$ is diagonal and nonsingular. 
Then (\ref{toda_1}) gives:
\BE 
  \omega_i(\vartheta) = \left\{ \begin{array}{ll}
   \frac{p_i\sqrt{\Delta_{ii}}}{2\sqrt{ \kappa_i}}
   \sin (\sqrt{ \kappa_i}(\vartheta-\theta_i)) 
 & \mbox{for   } \kappa_i>0, \\
   \frac{p_i\sqrt{\Delta_{ii}}}{2}
  (\vartheta-\theta_i)  & \mbox{for   } \kappa_i=0, \\
   \frac{p_i\sqrt{\Delta_{ii}}}{2\sqrt{-\kappa_i}}
\sinh(\sqrt{-\kappa_i}(\vartheta-\theta_i)) & \mbox{for   } \kappa_i<0,
  \end{array} \right.  
\label{om_sol_1}
\EE
where real phases $\theta_i$ are independent but (\ref{eq_en}) gives a 
restriction on $\kappa_i$:
\BE 
\sum_i \frac{\kappa_i}{\Delta_{ii}} = \frac12 (\Lambda_\chi + \Lambda_c). 
\EE
Substituting (\ref{om_sol_1}) into (\ref{sol_11}--\ref{sol_13}) and 
solving (\ref{sol_14}):
\BE 
\exp(C_i(\vartheta)) = \left\{ \begin{array}{ll}
   E_i-\frac{4\sqrt{ \kappa_i}}{p_i\Delta_{ii}}
\cot (\sqrt{ \kappa_i}(\vartheta-\theta_i)) & \mbox{for   } \kappa_i>0, \\
   E_i-\frac{4}{p_i\Delta_{ii}} 
(\vartheta-\theta_i)^{-1}  & \mbox{for   } \kappa_i=0, \\
   E_i-\frac{4\sqrt{-\kappa_i}}{p_i\Delta_{ii}}
\coth(\sqrt{-\kappa_i}(\vartheta-\theta_i)) & \mbox{for   } \kappa_i<0,
  \end{array} \right. 
\label{ec_sol_1} 
\EE
where $E_i$ are integration constants, an explicit form of the solution is 
obtained.
If $N_A=1$ we recover the solution found in \cite{Zhou}.

Recall also $D=11$ supergravity elementary-ansatz solution preserving 
supersymmetry.
In this case the space-time interval is \cite{DS}:
\BE 
ds^2 = \left(1+\frac{k}{r^6}\right)^{-2/3} dx^{\mu^{\{1\}}} dx_{\mu^{\{1\}}} 
 + \left(1+\frac{k}{r^6}\right)^{1/3}(dr^2+r^2 d\Omega^2) 
\label{interval_3} 
\EE
where two assuptions were made.
First, the geometry of the solution tends to flat empty space when 
$r\rightarrow +\infty$. 
Second, $k>0$ to avoid singularity for positive $r$.
Metric tensor coefficients are well defined if $r>0$, but the solution can 
be extended beyond point $r=0$ onto area not covered by $r$ coordinate 
\cite{GT+DGT}.
Rewriting (\ref{interval_3}) with $\vartheta$ one obtains:
\BE 
ds^2= \left(12 k |\vartheta-\theta|\right)^{-2/3} 
  dx^{\mu^{\{1\}}} dx_{\mu^{\{1\}}} +
  \left(6k |\vartheta-\theta| \right)^{1/3} 
 (6|\vartheta|)^{-7/3}(d\vartheta^2 + (6\vartheta)^2 d\Omega^2),
 \label{interval_4}
\EE
where $k=-1/(1+12\theta)$ and $\theta<-1/12$.
Preserving supersymmetry condition is equivalent in this case to $R=0$ 
(harmonic gauge) and $c_{\{1\}}=c_B=0$ \cite{DS}.
Flat space condition fixes values of $E_{\{1\}}$ and $E_B$.
Positive $r$ from (\ref{interval_3}) corresponds to 
$\vartheta \in (-1/12, +\infty)$ in (\ref{interval_4}) and
$r=0$ coincides with $\vartheta=+\infty$.
One can calculate $R_{MN}R^{MN}$ finding that it is proportional to 
$|\vartheta/(\vartheta-\theta)|^{14/3}$ and at
$\vartheta=\pm\infty$ has a nonzero finite value.
This gives a hint that $\vartheta=+\infty $ and $\vartheta=-\infty$ can be 
identified and the extension of the
solution beyond $\vartheta=\pm \infty$ is described by 
$\vartheta\in (-\infty, \theta)$.
At $\vartheta=\theta$ where $R_{MN}R^{MN}$ grows to infinity a true 
singularity is located.

Having (\ref{om_sol_1}) one can extend the transformation introduced in 
(\ref{sym_1}) onto space of all parameters appearing in the solution:
\begin{multline}  
(\vartheta ;  R,s_\chi, p_i, \theta_i, c_I, c_B, c_\alpha,E_i,E_I,E_B,
 E_\alpha,\kappa_i) \rightarrow \\
 \rightarrow (-\vartheta;1/R,s_\chi,-p_i,-\theta_i,-c_I,-c_B,-c_\alpha,
 E_i,E_I,E_B,E_\alpha,\kappa_i) 
  \label{sym_3} 
\end{multline}
and check that the transformation is a symmetry of all fields making up 
the solution: $e^{A_I}, e^{B_\vartheta}, e^{\phi_\alpha}$ and 
$e^{C_i}$ (\ref{sol_11}--\ref{sol_13}, \ref{ec_sol_1}).
But $\chi$ function (\ref{chi1}) is not invariant under (\ref{sym_3}).
This results in particular with fact that the supersymmetric solution 
given by (\ref{interval_4}) has a non-supersymmetric partner which can be 
described by the same formula with $k=-1/(1-12\theta)$.

\section{Conclusions}

In this paper we have shown that the system of equations of motions for
gravity coupled to antisymmetric tensors and scalar fields reduces to
Toda-like equations. The assumption of harmonic gauge usually made in
connection with supersymmetry is not needed if an adequate redefinition of
radial coordinate $r$ is done. A duality in the set of solutions
$\vartheta \rightarrow - \vartheta$ (or equivalently $r \rightarrow 1/r$)
was noticed.

The set of assumptions made in this paper follows the analogous set 
usually made in supergravity: there are several
 scalar and antisymmetric fields coupled to gravity and 
the solutions are not necessarily supersymmetric. 
Under these assumptions the extremely complicated system of equations 
reduces to a well-known Toda-like system. 

\vspace{10mm} 
\noindent 
{\Large \bf Acknowledgments} 

\vspace{5mm}
\noindent
\\ 
Work supported partially by the European Grant
HPRN--CT--2000--00152 Supersymmetry and the Early Universe.



\begin{thebibliography}{Ref}

\bibitem{LP1}   H. Lu, C.N. Pope {\it p-brane taxonomy.} In {\it Trieste 1996, High energy physics and cosmology}
                340-384. {\tt hep-th/9702086} and references in it.

\bibitem{Stelle} K.S. Stelle {\it BPS Branes in Supergravity.} In {\it Trieste 1997, High energy physics and cosmology.}
                29-127. {\tt hep-th/9803116} and references in it.

\bibitem{IM}    V.D. Ivashchuk, V.N. Melnikov {\it Exact solutions in multidimensional gravity with antisymmetric
                forms.} Class. Quant. Grav. {\bf 18}, R1-R66 (2001). {\tt hep-th/0110274} and references in it.

\bibitem{DGHR}  A. Dabholkar, G.W. Gibbons, J.A. Harvey, F. Ruiz Ruiz {\it Superstrings and solitons.} 
                Nucl. Phys. {\bf B340}, 33-55 (1990). 

\bibitem{DS}    M.J. Duff, K.S. Stelle {\it Multimembrane solutions of $D=11$ supergravity.} Phys. Lett. {\bf B253}, 
                113-118 (1991).

\bibitem{LPX}   H. Lu, C.N. Pope, K.W. Xu {\it Liouville and Toda solutions of M theory.} 
                Mod. Phys. Lett. {\bf A11}, 1785-1796 (1996). {\tt hep-th/9604058}

\bibitem{IK}    V.D. Ivashchuk, S.W. Kim {\it Solutions with intersecting p-branes related to Toda chains.}
                J. Math. Phys. {\bf 41}, 444-460 (2000). {\tt hep-th/9907019}

\bibitem{CGM}   S. Cotsakis, V.R. Gavrilov, V.N. Melnikov {\it Integrable spherically symmetric p-brane models
                associated with Lie algebras.} Grav. Cosmol. {\bf 6}, 66-75 (2000). {\tt gr-qc/0004052}

\bibitem{Tollsten} A.K. Tollsten {\it String solutions to supergravity.} {\tt hep-th/9610176}

\bibitem{Zhou}  B. Zhou, C.-J. Zhu {\it The complete black brane solution in D-dimensional coupled gravity system.}
                {\tt hep-th/9905146} 

                B. Zhou, C.-J. Zhu {\it The complete brane solution in D-dimensional coupled gravity system.} 
                Commun. Theor. Phys. {\bf 32}, 173-176 (1999). {\tt hep-th/9904157}

\bibitem{Gibbons} G.W. Gibbons {\it Branes and bions} Class. Quant. Grav. {\bf 16}, 1471-1477 (1999). 
               {\tt hep-th/9803203}

\bibitem{AC}    M. Toda {\it Theory of nonlinear lattices.} Springer-Verlag 1981.
 
                M. Toda {\it Theory of nonlinear waves and solitons.} Kluwer Dordrecht 1989.

                M.J. Ablovitz, P.A. Clarkson {\it Solitons, nonlinear evolution equations and inverse scattering.}
                Cambridge University Press 1991. 

\bibitem{GT+DGT} G.W. Gibbons, P.K. Townsend {\it Vacuum interpolation in supergravity via super p-branes.}
                Phys. Rev. Lett. {\bf 71}, 3754-3757 (1993). {\tt hep-th/9307049}

                M.J. Duff, G.W. Gibbons, P.K. Townsend {\it Macroscopic superstrings as interpolating solitons.}
                Phys. Lett. {\bf B332}, 321-328 (1994). {\tt hep-th/9405124}

\end{thebibliography}
\end{document}